# Non-Poissonian ultrashort, nanoscale electron pulses

Sam Keramati[1,*], Will Brunner[1], T. J. Gay[1], Herman Batelaan[1,†]

[1]*Department of Physics and Astronomy, University of Nebraska-Lincoln, Lincoln, NE 68588, USA*

**The statistical character of electron beams used in current technologies, as described by a stream of particles, is random in nature. Using coincidence measurements of femtosecond pulsed electron pairs, we report the observation of sub-Poissonian electron statistics that are non-random due to two-electron Coulomb interactions, and that exhibit an anti-bunching signal of 1 part in 4. This advancement is a fundamental step towards realizing a strongly quantum degenerate electron beam needed for many applications, and in particular electron correlation spectroscopy.**

Particle correlation spectroscopy, developed towards the end of the last century, is being used in an ever-lengthening list of applications. Photon correlations can be used, e.g., for sizing particles such as biological molecules and in aeronautical velocimetry [1]. In heavy-ion collisions, two-particle correlation measurements reveal the femtometer-size geometry of the ion source [2], while identical-atom correlation techniques demonstrate the quantum statistical nature of atomic isotope distributions [3]. The famous work of Hanbury Brown and Twiss (HBT) [4,5] and its elucidation by Glauber [6] led to this use of identical particle quantum correlations. A key underlying idea in all these experiments is that without particle correlation it becomes harder to resolve structural detail as the particles get closer together, while with particle correlations, proximity of the interacting particles makes such details easier to resolve. Examples of this include the resolution of stellar radii using light [7] and the femtometer sizing of nuclear structure using pions [2].

For electron sources, we have not yet seen the development of correlation spectroscopy, although the recent studies of Kiesel et al. [8] and Kuwahara et al. [9] show the way forward. In these experiments, a reduction in the number of coincidences at a pair of detectors was observed for free electrons emanating from small sources. The nano- and micron-scale size of the sources ensured that electrons were close together in the direction transverse to the electron beam axis. Nonetheless, the continuous cold-field emission [8] or nanosecond pulsed photoemission [9] of electrons in these investigations made it unlikely that the electrons were close together in time, so that the (expectedly) small detected deviation from random statistical behavior was found not to exceed 1 part in 1000, limited by the finite detection-time resolution.

Introducing new classes of electron beams has led [10-12] and should continue to lead [13,14] to new applications. As non-random correlation signals are strongly dependent on electron-electron proximity, and thus on the source size and the pulse duration, we expect that correlation spectroscopy can be used as a complement to streak imaging [15, 16] in the accumulation mode. Additionally, just as "femtoscopy" using the HBT effect probes the femtometer scale for high-energy collisions [2], using pulsed, correlated electrons provides a novel route to exploring small-scale surface phenomena at ultrafast time scales.

A strong deviation from random, Poissonian statistics can herald quantum-optical effects [6,17,18]. The bosonic quantum nature of photon HBT-bunching, leading to super-Poissonian statistics, can be compared to the fermionic nature of electron HBT-like anti-bunching, leading to sub-Poissonian statistics. These experimental signatures of spin-statistical dual quantum effects can also be caused by classically describable causes. For light such a cause can be thermal intensity fluctuations [19], while for electrons the cause can be Coulomb repulsion. For the purpose of reaching the quantum statistical regime for electron beams, it is thus necessary to distinguish Coulomb pressure from Pauli blockade. Indeed, even though the anti-bunching signal in earlier work [8] was attributed to quantum degeneracy, later analyses indicated that Coulomb pressure may explain the observation [20-23]. The very recent observation of electron anti-bunching [9] is the next important step as it combines polarized electron sources in an electron microscope with coincidence detection. The claim in that work is that the electron polarization dependence of the anti-bunching observation is due to the HBT effect and not to polarization-dependent trajectories of the photoemitted electrons. The recent advent of sub-micron femtosecond laser driven electron sources of spin-polarized electrons [24] can further assist coincidence techniques to help unravel the classical and quantum contributions; unlike Pauli "forces", Coulomb forces are not spin dependent.

In the experiments reported here, we used a Ti:Sapphire femtosecond laser oscillator to photoemit ultrashort electron pulses from the apex of a tungsten nanotip (Fig. 1a). This ensured that the electrons in each pulse were close together both transversely and temporally. Such a source was introduced by Hommelhoff et al. [25] and exists today in several laboratories [26-29]. However, the non-random nature of the pulsed electrons, the key finding of this work, has not been demonstrated until now. The time-coincidence technique we have used to characterize the pulsed electron beam has made its non-statistical nature manifest; we observe departures from the expected Poissonian distribution of one part in four for the two-electron coincidence rate associated with a single pulse.

The output beam of the laser had a pulse duration of approximately 100 fs with 13.2 ns between laser pulses. An incident laser pulse can give rise to two-electron pulse emission. Their coincidence detection rate can be reduced by a repulsive Coulomb force between the electrons (Fig.1b) or by Pauli blockade of symmetric spin states (Fig.1c). Our estimated emission rate of $10^7$ s$^{-1}$ and a repetition rate of $10^8$ s$^{-1}$ imply that most laser pulses produce no electrons. In this case, the probability of producing a larger number of electrons per pulse is correspondingly smaller for a random distribution.

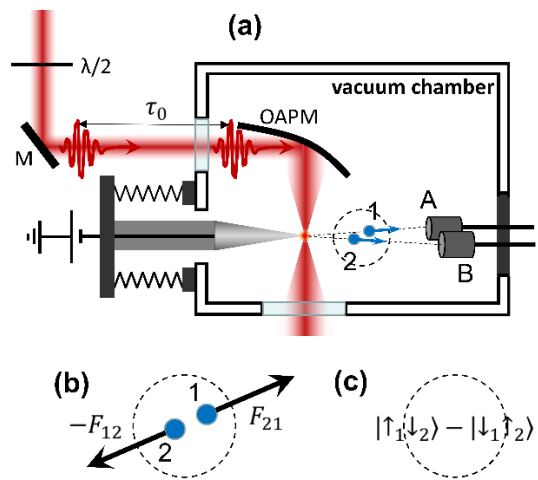

FIG. 1. Apparatus schematic. (a) Laser pulses (red) with a repetition time of $\tau_0$, tightly focused on an electrochemically etched tungsten nanotip (gray cone) [30] using an off-axis parabolic mirror (OAPM), induced emission of, e.g., single electrons or electron pairs (blue circular dots). The electrons were detected by two independent detectors (A, B) and the time delay $\tau$ between their arrivals was measured in coincidence (see text). The coincidence detection rate characterizes the presence of electron-electron interactions and a deviation from random electron arrival times. The coincidence spectra were obtained using NIM electronics [31]. Two examples of possible interactions are: (b) a repulsive Coulomb force that may reduce the coincidence rate or (c) a Pauli blockade that only allows a singlet state to populate a symmetrical orbital which may reduce the coincidence rate.

For our detection probability, about one detected electron is produced for every $10^5$ laser pulses, and a two-electron pulse occurs once for every ~$10^7$ laser pulses. Two electrons that arrive nearly simultaneously can be due to two electrons produced at the same time with nearly the same energy, but can also be due to two electrons produced at different times where the second electron has more energy and arrives at the same time as the first, lower-energy electron. This effect hampers experiments that are not time-resolved, but is significantly reduced for femtosecond pulsed sources when the pulses can be resolved in time.

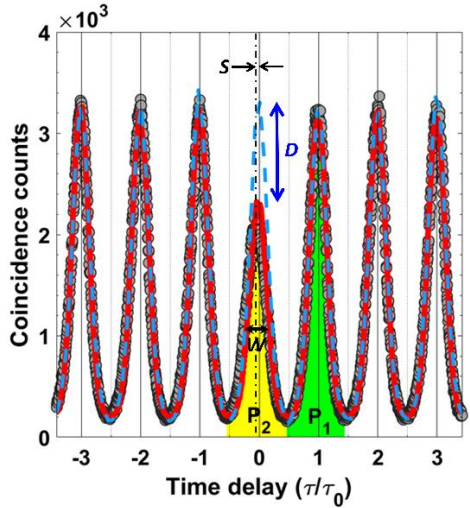

FIG. 2. Time delay coincidence spectrum. The experimental number of coincidence counts (circles) shows that the zero-time delay peak is smaller than its neighboring peaks. The zero-time delay peak contains two-electron pulses, while the neighboring peaks at integer multiples of the laser repetition time, $\tau_0 = 13.2\ ns$, contain single-electron pulses. A simulation of the experiment with a random number $n$ of electrons per pulse in the absence of electron-electron interactions predicts identical heights for all peaks (dashed blue line). The random character in the simulation is given by a Poissonian distribution $P_{\mathrm{n}}(\lambda)$, where $\lambda = r\tau_0$ and $r$ is the emission rate. A simulation using two-electron pulses that includes mutual Coulomb interaction (red solid line) predicts a reduced central peak consistent with the experiment. The coincidence dip $D$ is caused by the component of the mutual Coulomb force that is transverse to the electrons' motion. A small change in the width $W$ and a small shift $S$ of the zero-delay peak can also be attributed to the Coulomb force between the two electrons (see text).

The main experimental result is shown in Fig. 2, where the number of coincidences is shown as a function of the time delay between two detected electrons. The data acquisition time was 15 minutes. The peak at zero-delay time is due almost exclusively to two electrons generated by the same laser pulse. The peak at $\tau = \tau_0$ is due to coincidence events where an electron generated by a laser pulse triggers the "start" detector, followed by another electron generated by the next laser pulse that triggers the stop detector. For the $\tau = 2\tau_0$ peak, the next laser pulse does not give rise to a stop trigger but the second-to-next laser pulse does, and so forth. An electronic time delay placed in the stop channel ensures that peaks with a negative time delay can also be recorded. The main observation of this work is that the zero-delay time peak is reduced compared to the surrounding peaks. We refer to this reduction as the "dip," and its presence is a clean signature that the statistical nature of the electron emission process is non-Poissonian [32]. The dip does not depend on the details of the detectors' efficiency or asymmetry.

To explain the presence of the dip in the zero-time delay peak, we performed a simulation with Coulomb interaction between the two electrons [33]. The simulation result (red solid line) agrees with the observed dip of Fig. 2 (circles). The transverse component of the mutual Coulomb force tends to push some electrons outside of the effective detection area (Fig. 3a and 3b), thereby lowering the count in the zero-time delay peak. To find the height of the neighboring peaks, the mutual Coulomb interaction is turned off. By varying the strength of the Coulomb interaction in the simulation and inspecting the electron trajectories we find that the dip value $D$ first increases as more electron pairs are pushed into the detectors (moving from region 1 to 2 in Fig. 3a and 3b), before it decreases when electrons are pushed outside the detection range (region 3 in Fig. 3a and 3b). The value of $D$ increases with a decreasing time interval, $\Delta t_e$ [33], between the electrons' emission times. For Gaussian shaped laser pulses, having intensity $I(t)$ with a FWHM of 100 fs, the electron pulse duration is 50 fs for an $I^4(t)$ (4-photon photoemission) process [34]. Such an estimate is valid for single-electron pulses. For the two-electron pulses recorded in the zero-delay peak, an intensity dependence $\propto I^8$ is expected. For Gaussian temporal envelopes, the standard deviation of $\tau$ is 30 fs, that is, about 68% of the electron pairs are emitted with a temporal separation of less than 30 fs. A simulated pulse duration $\Delta t_e$ of 10 fs yields a dip size that agrees with the experimental data. The tip radius $R_{\text{tip}}$ was taken to be 25 nm. This is about half the value observed by SEM. The "lightning rod" effect [35] and the detector geometry, which selects electrons emitted in the forward direction, likely explains emission from a smaller size area. The main reported effect, that is, the reduction of the central peak, $D$, is robust and was observed for different tungsten nanotips.

We now show that Pauli blockade does not contribute to the dip. The Hanbury Brown-Twiss relative dip size due to Pauli blockade $D_{\text{HBT}}$ ($D_{\text{HBT}} = 0$ for no dip; $D_{\text{HBT}} = -1$ corresponds to a complete dip) was studied in detail for 1D propagation [23]. For unpolarized electrons, straightforward generalization to 3D propagation yields the estimate

$$D_{\text{HBT}} = -\frac{1}{2} \frac{\tau_c}{\Delta t_e} \frac{X_c}{X_{\text{tip}}} \frac{Y_c}{Y_{\text{tip}}}, \tag{1}$$

in which $\tau_c \leq \Delta t_e$, $X_c \leq X_{\text{tip}}$, and $Y_c \leq Y_{\text{tip}}$ are the coherence time and lengths and $X_{\text{tip}} \sim Y_{\text{tip}}$ are the physical sizes of the tip. The minimum energy-time uncertainty relation limits the coherence time to $\tau_c = \hbar/2\Delta E = 0.66\ fs$ for $\Delta E = 0.5\ eV$; the coherence lengths are estimated from the position-momentum uncertainty relation as follows. The angular width of the distribution of the electrons that hit the detectors is $\gamma = \tan^{-1}\left(\frac{\Delta_{wx} + X_{det}}{L}\right)$, where $X_{det}$ is the distance between the centers of the front openings of the detectors, $\Delta_{wx}$ is the opening size of the detectors and $L$ is the distance from the tip to the detectors [33].

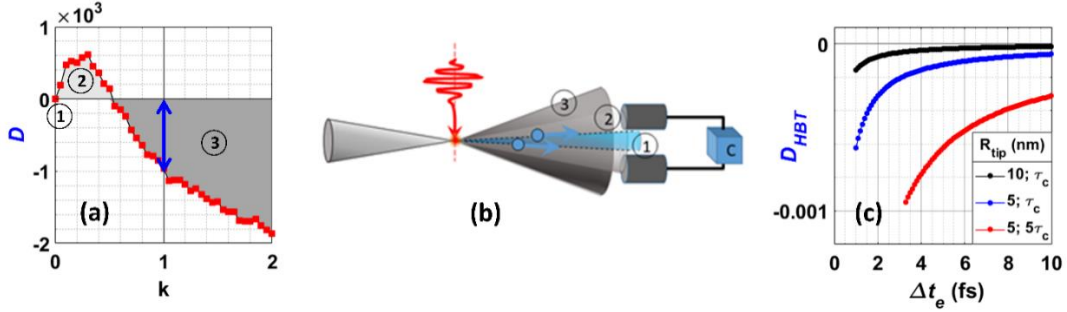

FIG. 3. Coulomb model. (a) The strength of the mutual Coulomb interaction k is varied in the model with k=1 corresponding to that which occurs in nature. In the absence of Coulomb interaction (region (1)), some forward propagating electron pairs miss the detectors. With increasing k, these electrons are pushed into the detectors increasing the number of coincidence counts $D$ (region (2)). In region (3), the Coulomb interaction is strong enough to push electron pairs out of the detection range so that $D$ becomes negative. (b) The three trajectory regions described in (a) are illustrated schematically with three emission cones. (c) Pauli blockade due to quantum degeneracy is estimated to lead to a 0.01 – 0.1% reduction of the normalized coincidence count rate and thus cannot explain the observed 24% reduction. This conclusion is not changed when considering a significantly smaller tip apex radii, $R_{tip}$, a factor of two shorter pulse duration, and an extended coherence time of $5\tau_c$.

The transverse ($x$) component of the linear momentum uncertainty is $\Delta p_x = p\gamma$, where $p = m_e v$. The transverse coherence lengths are given by $X_c = \hbar/2\Delta p_x$ and $Y_c = \hbar/2\Delta p_y$, where $\Delta p_y = p\tan^{-1}\left(\frac{\Delta_{wy}}{L}\right)$. The emission site lateral widths are estimated by $X_{tip} = 2R_{tip}\sin\gamma$, and $Y_{tip} = X_{tip}$. For these parameters the contribution of quantum degeneracy to anti-bunching is less than 1 part in 1000 (Fig. 3c).

An example of electron correlation spectroscopy is the use of the method described here to obtain the approximate duration of the electron pulse. To do this, the simulation model used to calculate the temporal spectrum (Fig. 2, red solid line) is simplified by calculating the impulse imparted to the electron along straight trajectories. This perturbative approach is motivated both by the result from the full simulation that trajectory deflection due to the mutual Coulomb forces is small, and the long computation times required for a full simulation. The result is that at large temporal separations the dip becomes negligible for all nanotip radii [36], substantiating the claim that an upper limit to the pulse duration can be estimated, and that for the experiment reported here the pulse is shorter than 10 fs.

Finally, we consider the pathway to produce electron beams with strong degeneracy. In our work, the transverse coherence can be increased by a diverging electron lens [8]. For a 100X magnifying lens, the detected transverse electron momentum is decreased and $X_c = \hbar/2\Delta p_x$ increased to reach full transverse coherence. In this case Eq. 1 becomes $D_{HBT} = -\frac{1}{2}\frac{\tau_c}{\Delta t}$, where $\Delta t$ is either $t_{pulse}$ or $t_{detector}$, depending on whether the pulses are resolved in time or not. In earlier work [8,9,21] the longitudinal coherence was low which explains why the observed dips were limited to about 1 part in 1000, as the dip appears in the second order correlation function for unpolarized electrons as $g^{(2)} = 1 - t_c/2\Delta t$. The addition of temporal resolution may thus lead to strong degeneracy. Note that the expression is the same as that for pulsed X-rays, for which $g^{(2)} = 1 - t_c/2t_{pulse}$ [37]. This has been used as a method for determining X-ray pulse duration [38] and constitutes an example of photon correlation imaging. In our work the ratio of $t_{coherence}$ to $t_{pulse}$ is about $10^{-1}$ and can be increased using an energy analyzer to reduce the energy spread of the detected electrons. An analyzing power of 1-100 meV corresponds to $T_{coh}$ of ~100-1 fs. Analyzing powers reaching sub-10 meV have been realized [39], indicating that the observation of strong quantum degeneracy in an electron beam is, in principle, within reach [40]. The use of a femtosecond source could thus help reach the quantum degenerate regime. This was considered before [23, 41] but it led to a lower degeneracy [41]. Nevertheless, when the pulses are temporally resolved (Fig. 2a), the degree of degeneracy of the source is not necessarily a deciding factor if the goal is to observe a time-resolvable coincidence dip. Instead, the capability to compare the number of two-electron-events with single electron-events by post-selection becomes important.

In summary, we demonstrated that a tungsten nanotip electron source driven by femtosecond laser pulses exhibits a strong deviation from random Poissonian statistics. Because the pulses are time resolved, multi-electron pulses are distinguished from single-electron pulses even when the emission rate is low. Coulomb interactions explain our data and are sensitive to the tip apex radius and the electron pulse duration. As a result, this technique provides a method to characterize ultrashort electron pulse duration and photoemitting nanoscale structure size. The results also indicate that in HBT-type experiments with free electrons [8], the mutual Coulomb repulsion can contribute to the anti-bunching signal, and should not be neglected as it was in earlier work, as pointed out in refs. [20, 21, 22]. A larger source size and a longer pulse length can suppress the Coulomb interaction [9], while a diverging lens [8] and an energy analyzer can enhance the Pauli blockade. A laser-driven spin polarized source [9, 24] can serve as a means to distinguish their relative contribution for femtosecond pulses [22, 23]. Femtosecond sources and time-resolved correlation techniques are expected to provide strong signals in the quantum regime. This provides access to new ideas in quantum statistics and the Pauli exclusion principle [42, 43, 44], leads to techniques that benefit from heralded single-electron on-demand sources such as quantum electron microscopy [14] and electron ghost imaging [45], and helps develop entanglement-assisted [46] electron microscopy [47, 48] as well as ultrafast electron microscopy [49, 50].

S. Keramati and H. Batelaan acknowledge support for this work by the National Science Foundation (NSF) under the award number PHY-1912504, and the Nebraska Research Initiative. S. Keramati, W. Brunner and T. J. Gay acknowledge support by NSF under the award number PHY-1806771. The SEM images were taken at the NanoEngineering Research Core Facility (NERCF), which is partially funded by the Nebraska Research Initiative.

*sam.keramati@huskers.unl.edu

†hbatelaan@unl.edu

1. *Light Scattering and Photon Correlation Spectroscopy*, E.R.Pike and J.B. Abbiss, eds., NATO ASI Series 3. High Technology, Vol. 40 (Springer Science+Business Media, B.V., 1997)

2. Baym, G. The physics of Hanbury Brown–Twiss intensity interferometry: from stars to nuclear collisions. *Acta. Phys. Pol.* B **29**, 1839–1884 (1998).

3. Jeltes, T., et al. Comparison of the Hanbury Brown–Twiss effect for bosons and fermions, *Nature* **445**, 402–405 (2007).

4. Hanbury Brown, R. & Twiss, R.Q. A new type of interferometer for use in radio astronomy. *Philosophical Magazine* **45**, 663–682 (1954).

5. Hanbury Brown, R. & Twiss, R. Q. Correlation between Photons in two Coherent Beams of Light. *Nature* **177**, 27–29 (1956).

6. Glauber, R. J. *Quantum Theory of Optical Coherence. Selected Papers and Lectures* (Wiley-VCH, Weinheim, 2007).

7. Hanbury Brown, R. & Twiss, R.Q. A Test of A New Type Of Stellar Interferometer On Sirius. *Nature* **178,** 1046–1048 (1956).

8. Kiesel, H., Renz, A. & Hasselbach, F. Observation of Hanbury Brown-Twiss anticorrelations for free electrons. *Nature* **418**, 392-394 (2002).

9. Kuwahara M., et al. Intensity Interference in a Coherent Spin-Polarized Electron Beam, *Phys. Rev. Lett.* **126**, 125501 (2021).

10. Zewail, A. H. Four-dimensional electron microscopy. *Science* **328**, 187-193 (2010).

11. Miller, R. J. D. Femtosecond crystallography with ultrabright electrons and X-rays: capturing chemistry in action. *Science* **343**, 1108-1116 (2014).

12. Arbouet, A., Caruso, G. M. & Houdellier, F. Ultrafast transmission electron microscopy: historical development, instrumentation, and application. *Advances in Imaging and Electron Physics* Vol. 207, pp 1-72. (Elsevier, Amsterdam, 2018).

13. Hassan, M. Th. Attomicroscopy: from femtosecond to attosecond electron microscopy. *J. Phys. B: At. Mol. Opt. Phys.* **51**, 032005 (2018).

14. Kruit, P. Designs for a quantum electron microscope. *Ultramicrosocopy* **164**, 31-45 (2016).

15. Liang, J., Zhu, L. & Wang, L.V. Single-shot real-time femtosecond imaging of temporal focusing. *Light Sci. Appl.* **7,** 42 (2018).

16. Gao, M. et al., Full characterization of RF compressed femtosecond electron pulses using ponderomotive scattering, *Optics Express* **20**, 12048-12058 (2012).

17. Hong, C. K., Ou, Z. Y. & Mandel, L. Measurement of subpicosecond time intervals between two photons by interference. *Phys. Rev. Lett.* **59** 2044–2046 (1987).

18. Jones, E., Becker, M., Luiten, J. & Batelaan, H. Laser control of electron matter waves, *Laser Photonics Rev.* **10**:2 214–229 (2016).

19. Loudon, R., Photon Bunching and Antibunching. *Phys. Bull.* **27** 21-23 (1976).

20. Baym, G. & Shen, K., *In Memory of Akira Tonomura*, Edited By: Kazuo Fujikawa and Yoshimasa A Ono. Hanbury Brown–Twiss Interferometry with Electrons: Coulomb vs. Quantum Statistics. 201-210 (World Scientific, Singapore, 2014).

21. Kodama, T., Osakabe, N. & Tonomura, A. Correlation in a coherent electron beam. *Phys. Rev. A* **83**, 063616 (2011).

22. Lougovski, P. & Batelaan, H. Quantum description and properties of electrons emitted from pulsed nanotip electron sources. *Phys. Rev. A.* **84**, 023417 (2011).

23. Keramati, S., Jones, E., Armstrong, J. & Batelaan, H. Partially coherent quantum degenerate electron matter waves. *Advances in Imaging and Electron Physics*. **213**, 3-26. (Elsevier, Amsterdam, 2020).

24. Brunkow, E., Jones, E. R., Batelaan, H. & Gay, T. J. Femtosecond-laser-induced spin-polarized electron emission from a GaAs tip, *Appl. Phys. Lett.* **114**, 073502 (2019).

25. Hommelhoff, P., Sortais, Y., Aghajani-Talesh, A. & Kasevich, M. A. Field Emission Tip as a Nanometer Source of Free Electron Femtosecond Pulses. *Phys. Rev. Lett.* **96**, 077401 (2006).

26. Ropers, C., Solli, D. R., Schulz, C. P., Lienau, C. & Elsaesser, T. Localized Multiphoton Emission of Femtosecond Electron Pulses from Metal Nanotips. *Phys. Rev. Lett.* **98**, 043907 (2007).

27. Barwick, B. et al. Laser-induced ultrafast electron emission from a field emission tip. *New J. Phys.* **9**, 142 (2007).

28. Yanagisawa, H., Optical Control of Field-Emission Sites by Femtosecond Laser Pulses. *Phys. Rev. Lett.* **103**, 257603 (2009).

29. Vogelsang, J. Ultrafast Electron Emission from a Sharp Metal Nanotaper Driven by Adiabatic Nanofocusing of Surface Plasmons. *Nano Lett.* **15**, 4685 (2015).

30. See Supplemental Material at [URL will be inserted by publisher] for the nanotip manufacture procedure.

31. See Supplemental Material at [URL will be inserted by publisher] for coincidence circuitry.

32. See Supplemental Material at [URL will be inserted by publisher] for statistical analysis proofs.

33. See Supplemental Material at [URL will be inserted by publisher] for numerical simulation details.

34. Hilbert, S. A., Barwick, B., Fabrikant, M., Uiterwaal, C. J. G. J. & H. Batelaan. A high repetition rate time-of-flight electron energy analyzer. *Appl. Phys. Lett.* **91**, 173506 (2007).

35. Batelaan H. & Uiterwaal, C. J. G. J. Nature **446**, 500 (2007).

36. See Supplemental Material at [URL will be inserted by publisher] for temporal limits.

37. Classen, A., Ayyer, K., Chapman, H. N., Röhlsberger, R. & von Zanthier, J. Incoherent Diffractive Imaging via Intensity Correlations of Hard X Rays, *Phys. Rev. Lett.* **119**, 053401 (2017).

38. Yabashi, M., Tamasaku, K. & Ishikawa, T. Measurement of X-Ray Pulse Widths by Intensity Interferometry, *Phys. Rev. Lett.* **88**, 244801 (2002).

39. Allan, M. Threshold Phenomena in Electron-Molecule Scattering. *Phys. Scr.* **T110**, 161 (2004).

40. See Supplemental Material at [URL will be inserted by publisher] for a quantum degeneracy parameter map.

41. Kuwahara, M., et al., The Boersch effect in a picosecond pulsed electron beam emitted from a semiconductor photocathode. *Appl. Phys. Lett.* **109**, 013108 (2016).

42. Silverman, M. P., On the feasibility of observing electron antibunching in a field-emission beam, *Phys. Lett*. A **120**, 442 (1987).

43. Kodama, T. et al., Feasibility of observing two-electron interference, *Phys. Rev. A* **57**, 2781 (1998).

44. *Spin-Statistics connection and the commutation relations.*, Editors: Robert C. Hilborn, Gugliemo M. Tino (American Institute of Physics, Melville NY, 2000).

45. Li, S. et al. Electron Ghost Imaging, *Phys. Rev. Lett.* **121**, 114801 (2018).

46. Schattschneider, P. & Löffler, S. Entanglement and decoherence in electron microscopy. *Ultramicroscopy* **190**, 39–44 (2018).

47. Okamoto, H. & Nagatani, Y. Entanglement-assisted electron microscopy based on a flux qubit. *Appl. Phys. Lett.* **104**, 062604 (2014).

48. Okamoto, H. Possible use of a Cooper-pair box for low-dose electron microscopy. *Phys. Rev. A*, **85**, 043810 (2012).

49. Priebe, K. E. et al. Attosecond electron pulse trains and quantum state reconstruction in ultrafast transmission electron microscopy. *Nat. Photonics* **11**, 793-797 (2017).

50. Morimoto, Y. & Baum, P. Diffraction and microscopy with attosecond electron pulse trains. *Nat. Phys.* **14**, 252-256 (2018).

# Supplementary Material

# Non-Poissonian ultrashort, nanoscale electron pulses

Sam Keramati[1,*], Will Brunner[1], T. J. Gay[1], Herman Batelaan[1]

[1]Department of Physics and Astronomy, University of Nebraska-Lincoln, Lincoln, NE 68588, USA

**TUNGSTEN TIP MANUFACTURE**

The nanotip needle sources were fabricated by electrochemical etching of a tungsten wire with thickness 200 μm in a KOH solution through the "lamella drop-off" method [1] (Fig. S1). The experiments were performed in a vacuum chamber at a pressure of approximately $2\times10^{-7}$ Torr. At this pressure the effective radius of the nanotips can change while in the laser focus. To monitor this, the tip bias potential corresponding to the onset of DC field emission tunneling was checked regularly. The voltage was varied between about -250 V to -350 V during all the experiments indicating little nanotip reformation. We used a femtosecond laser with a repetition rate of 76 MHz, a spectral distribution centered at 800 nm, and a pulse energy of 1.6 nJ, corresponding to a peak intensity $I_0 \sim 5\times10^{14}\, Wm^{-2}$ for an estimated focal spot radius of 3 μm after reflection from a gold-coated off-axis parabolic mirror. The dominant electron emission involved a non-linear $I^4$ process indicative of four-photon emission.

**COINCIDENCE ELECTRONICS**

A pair of Dr. Sjuts Model KBL-510 Channel Electron Multipliers (CEMs) were used for electron detection (Fig. S1). The output pulses of the detectors were amplified with ORTEC VT-120 low-noise pre-amplifiers, and subsequently triggered an ORTEC model 935 constant fraction discriminator to indicate an electron detection. The count rates of both detectors were recorded using a Multi-Channel Scaler (MCS). An electron detection at one detector started, while a detection at the other detector stopped an ORTEC model 567 Time-to-Amplitude-Converter (TAC). An ORTEC model 425A adjustable delay module was used in the stop-signal channel. A coincidence count histogram of the TAC output was generated by an Multi-Channel-Analyzer (MCA) with 2048 digital channels. The coincidence window is 90 ns and the histogram bin width is 53.0 ps.

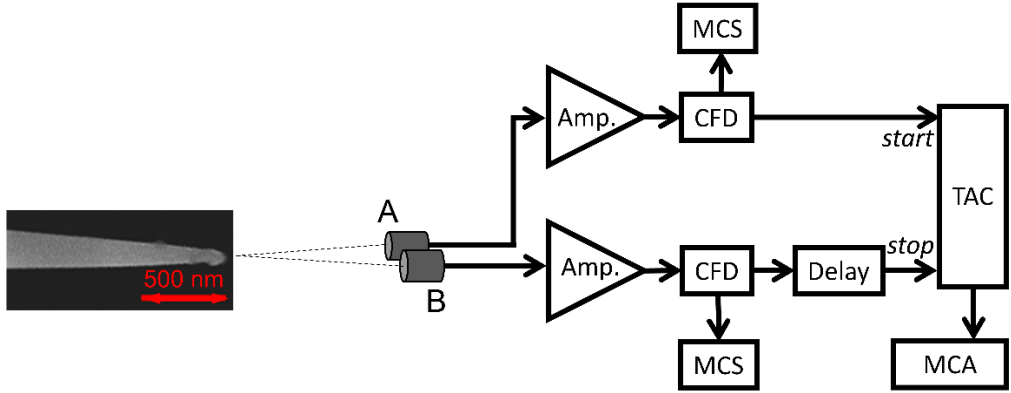

FIG. S1. Circuit schematic. Laser pulses with a repetition time of $\tau_0$, tightly focused onto an electrochemically etched tungsten nanotip (SEM image) induce emission of electron pairs (dashed lines). The electrons are detected by two independent detectors (A, B) and the time delay $\tau$ between their arrivals is measured in coincidence (see main text). The coincidence spectra are obtained using NIM electronics including low-noise Amplifiers (Amp.), Constant-Fraction Discriminators (CFD), Multi-Channel Scalers (MCS), a Time-to-Amplitude Converter (TAC), and a Multi-Channel Analyzer (MCA).

**MCA CALIBRATION**

The MCA channel (bin) widths for the coincidence window was determined using a pulse generator and adjustable calibrated external delay. The zero-time delay was found as follows. The six neighboring peaks of the zero-time delay peak are dominated by single-electron events and thus do not undergo a lateral shift as discussed in the main text. Counting the peaks from left to right, the 4$^{th}$ one is the zero-delay peak. By fitting a line to a plot of "peak number" vs. "MCA channel number" using only the 6 tall peaks, the time delay corresponding to the 4$^{th}$ peak was inferred.

**SIMION SIMULATION**

Individual electron trajectories were simulated with an energy of 100 eV and an energy spread of 1 eV using SIMION [2]. They were emitted isotropically from a point source in a solid cone corresponding to the circular aperture preceding the detectors. The detector openings were modeled as solid rectangles. The program simulated an experiment of 70 s duration and recorded each detection time. An electron pulse lasting 50 fs was generated every $\tau_0$ = 13.2 ns. The number of electrons per pulse was determined by Eq. (1) with an emission rate in the cone of r = 7.71×10$^5$ s$^{-1}$. Pulses that contained up to three electrons were taken into account. The electrons were emitted randomly during the 50 fs pulse duration. The raw data were subsequently analyzed with a MATLAB code that generated the coincidence histograms from the set of detection times at the two detectors.

The neighboring peaks (main text Fig. 2) have a width of approximately 5 ns, which is attributed to the finite detector resolution. Coincident time spectra where a laser pulse generates a start pulse and an electron generates a stop pulse yield peaks with a width of about 3 ns, characterizing the time resolution of the electron detector. For the electron-electron coincidence spectra the peaks are broadened by the combined resolution of two electron detectors. A set of 4 Gaussian random number generators with different widths were used in the MATLAB code to match the coincidence peaks in the experimental data.

**COULOMB INTERACTION SIMULATION**

To explain the presence of the "dip", the mutual electron-electron Coulomb interaction was simulated by integrating the equations of motion for the two interacting particles. Not only the dip, but two more parameters characterize small changes in the zero-time delay peak (main text Fig. 2). It is shifted by an amount *S* and has a different width *W* than its neighboring peaks. The simulation result (red solid line) agrees with the observed dip and provides a qualitative description of the less pronounced width change and peak shift. In this simulation, pairs of electrons were considered, and their motion was found by solving Newton's equations:

$$\begin{cases} m_e \, d^2\vec{r}_1/dt^2 = \vec{F}_{tip,1} + \vec{F}_{1,2} + \vec{F}_{det,1} \\ m_e \, d^2\vec{r}_2/dt^2 = \vec{F}_{tip,2} + \vec{F}_{2,1} + \vec{F}_{det,2} \end{cases}, \qquad (S1)$$

where the leading electron is labeled 1 and the trailing electron is labeled 2, $m_e$ is the electron mass, $\left|\vec{F}_{1,2}\right| = \left|\vec{F}_{2,1}\right| = F_{qq}$ is the mutual electron interaction, and $\vec{F}_{det,1}$ ($\vec{F}_{det,2}$) is the Coulomb force exerted on electron *1* (*2*) by the detector entrance biased at $V_f$. These equations were solved numerically using the Runge-Kutta method.

The nanotip apex is modeled as a hemisphere with radius $R_{tip}$. The electron's initial positions are chosen randomly according to a uniform distribution on the hemisphere within a cone around the z-axis. The opening angle of the cone is chosen to correspond to the physical aperture in front of the detectors. The initial angle with respect to the normal of the hemisphere's surface follows a random distribution with a cosine distribution (that is, peaked along the normal with zero emission probability parallel to the surface). The azimuthal angle around the normal is distributed uniformly. The initial electron energy is 0.5 eV and most of the acceleration due to the nanotip's electric field occurs in the first few hundred nanometers to an approximate energy of 100 eV. (Energy widths typically associated with multiphoton electron emission from tungsten do not significantly change the results of these simulations.) The accelerating force is indicated with $F_{tip}$. The second electron launches at a random time after the first electron within a time window $\Delta t_e$ (10 fs for the result in Fig. 2 main text).

The trajectories of the two electrons are described by the time-dependent position vectors labeled $r_1$ and $r_2$, at angles $\theta_1$ and $\theta_2$ with respect to the z-axis, respectively, in Fig. S2 (not shown are the azimuthal angles of the electrons). The e⁻ - e⁻ distance is $r_{qq}$ and the mutual Coulomb force is $F_{qq}$. The effect of the force on an emitted electron by the ionized tip is negligible as compared to the mutual Coulomb force.

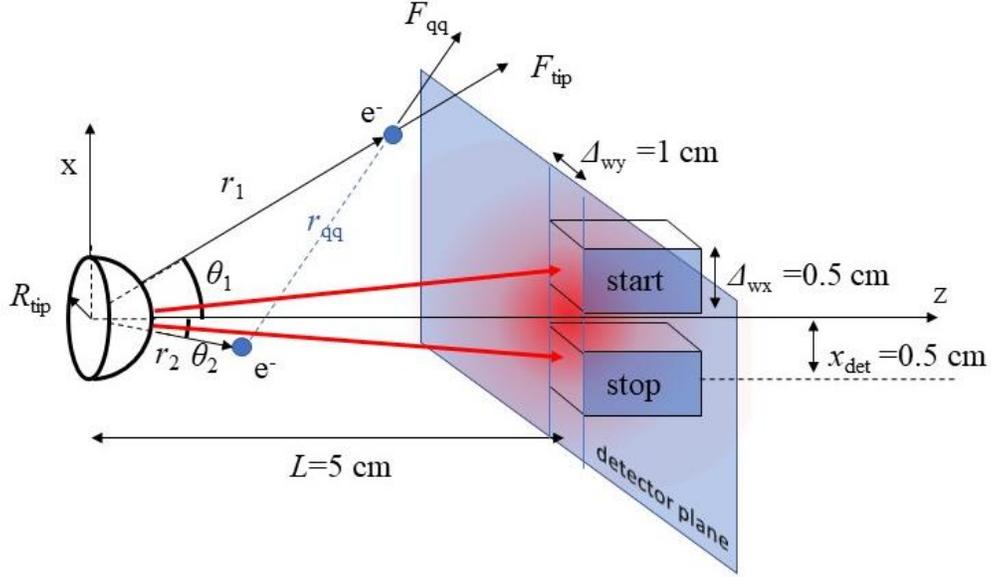

FIG. S2. Coulomb model schematic. A hemisphere with radius $R_{tip}$ is used to model the nanotip apex. Pairs of mutually interacting electrons propagating from the hemisphere to the detectors are considered. The arrival time delay is recorded. The definition and the values of the parameters are labeled in the figure and described in the text.

Just before the electrons reach the detection plane, they are slowed over a range of 5 mm by a repulsive force due to the negative voltage placed on the front of the detectors that is used to repel lower energy secondary electrons. The x-y detection plane is $L = 5$ cm away from the nanotip. The center of a circular aperture of diameter 1.4 cm is located on the z-axis (in the x-y plane, not shown in Fig. S2) and at a distance of 3.4 cm from the nanotip. This aperture is used in the experiment to suppress background electrons that scatter from the vacuum chamber walls. The rectangular opening faces of the start and stop CEM pair have sides of length $\Delta_{wx}$ and $\Delta_{wy}$. The distance from the center of each detector face to the z-axis is denoted by $x_{det}$. Most of the electron pairs propagate nominally in the forward direction illustrated by the red arrows in Fig. S2. When one electron passes the detection plane into the start detector and the other into the stop detector the event's time delay is recorded. A random Gaussian detection time is included to model the detector response.

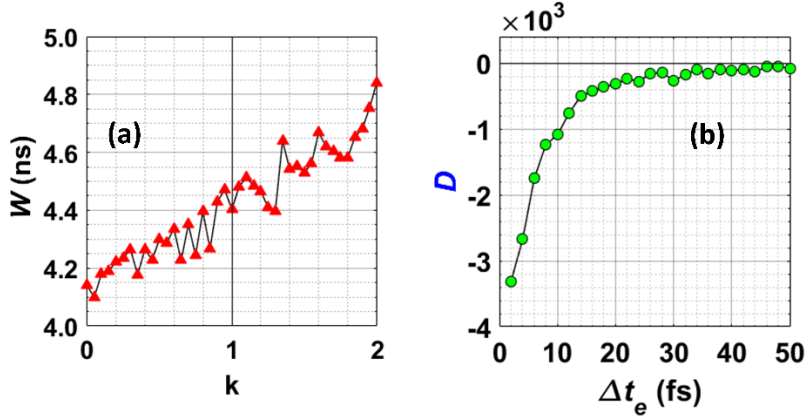

FIG. S3. Coulomb model results. (a) The strength of the mutual Coulomb interaction k is varied in the model. At k=1 the strength is that which occurs in nature. The width $W$ of the zero-time delay peak increases with k due to longitudinal Coulomb forces. (b) The dip increases for shorter electron pulse durations $\Delta t_e$ due to transverse Coulomb forces (for a fixed tip apex radius of 25 nm).

The width of the observed zero-time delay peak is slightly broader than that of the taller peaks by 0.8 ns (main text Fig. 2). This is qualitatively explained by the longitudinal component of the mutual Coulomb force that tends to increase the temporal separation. The Coulomb model results presented in Fig. S3a correspond to a broadening of 0.2 ns. Reducing the pulse length increases this value, but also increases the dip (Fig. S3b). The experimental shift $S$ of the zero-delay peak is ~0.5 ns, but the simulation result shown in Fig. 2 of the main text yields no shift. However, when the electron beam in the simulation is shifted laterally by a few millimeters, the resulting detection asymmetry does yield a shift of this size. Systematic experimental exploration of the broadening and shift was hampered by spurious electromagnetic fields that are sufficient to cause deflection of the electron beam by several millimeters.

**PULSE DURATION ANALYSIS**

The results of the simplified model using the perturbative approach are described in the main text. In this model, the accelerating trajectory of the first electron starting with an initial velocity, $v_{in}$, orthogonal to the nanotip surface with radius $R_{tip}$, is computed by numerical integration. The initial position of the first electron is determined by angles $\theta$ and $\varphi$ with respect to the nanotip symmetry axis. The second electron is launched at an angle $\theta$ and $\varphi+\pi$ after a time $\Delta t$. The arrival locations at the detection plane are tested to record a coincidence event. The ratio of the number of coincidence events with the mutual Coulomb repulsion to the number of events without the mutual Coulomb repulsion determines the "dip" size, here indicated as $D^{rel}$ for the simplified model.

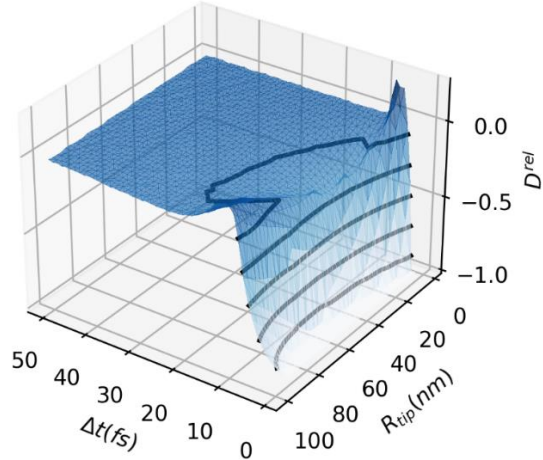

FIG. S4. Temporal limits. The relative dip size of the simulated coincidence signal is shown as a function of the time $\Delta t$ between the emission times of two electrons and the nanotip radius $R_{tip}$. Black lines are contours of constant dip size of -0.1, -0.3, -0.5, -0.7, and -0.9. A dip value of -0.3 indicates that the electrons have a maximum temporal separation of about 10 fs. This illustrates that an anti-bunching dip can be used to estimate the electron pulse duration.

In the 25 to 75 nm size range at a temporal separation of less than 10 fs, the Coulomb repulsion and dip size are appreciable (Fig. S4). The "dip" can also become a "peak" when the electron pairs experience weak Coulomb repulsion. Pairs that start at small $\theta$ values between the two detectors are pushed into the detector (also see Fig. 3b in the main text), while electrons that start at $\theta$ values which were aimed at the detectors experience almost no Coulomb repulsion and do not compensate for the increase in coincidence rate. At small tip radii both electrons are initially located close to the symmetry axis and the Coulomb force is directed mostly in the longitudinal direction. The small lateral component of the Coulomb force outweighs their close proximity and the effect of the Coulomb interaction on the measured coincidence signal is weak. At very large tip radii (not shown in the figure) the larger electron-electron separation reduces the Coulomb force effects even if the force is mostly lateral.

**QUANTUM DEGENERACY MAP**

The dip size due to the quantum degeneracy as estimated from Eq. 2 is plotted using a color map (Fig. S5). Blue indicates a strong quantum anti-bunching effect. The experiment of Kiesel et al. [3] reports an estimated coherence time of 32.5 fs, a detector resolution of 26 ps, and a dip of $1.26\times10^{-3}$; Kodama et al. [4] report an estimated coherence time of 3.9 fs, a detection resolution of 20 ps, and a dip of $1.1\times10^{-4}$; Kuwahara et al. [5] report an estimated coherence time of 90 fs, a detection resolution of about 170 ps, and a dip of about $10^{-3}$. Three data points are plotted outside of the figure frame to emphasize that other parameters for these experiments such as those associated with the electron source are not comparable with those of the work reported here. These parameters influence the effect of the classical Coulomb repulsion. The red color indicates a dominant classical anti-bunching effect, while in the purple region both Coulomb repulsion and the Pauli blockade play a strong role. We report an estimated coherence time of 0.66 fs, temporal width of 10 fs, and a dip of 0.25 (data point "[this work]" in Fig. S5). The color map is computed for a 100X magnifying lens for full transverse coherence.

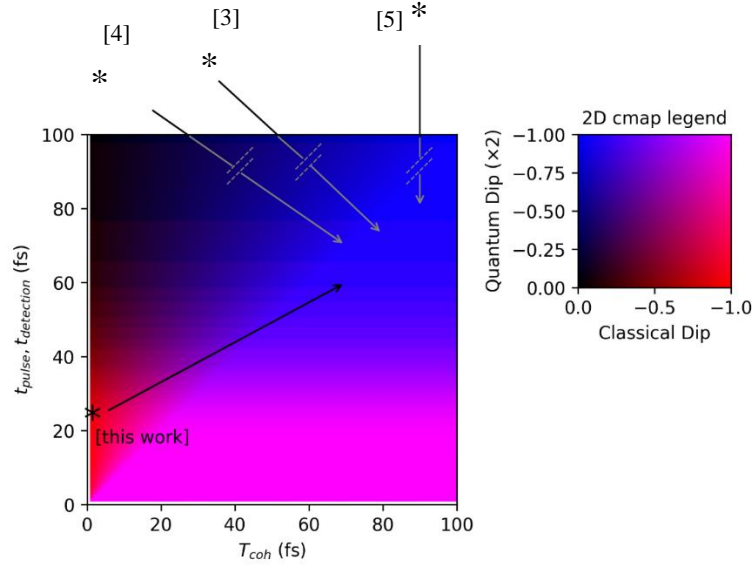

FIG. S5. Quantum degeneracy and Coulomb interaction. The quantum and classical contributions to the simulated coincidence dip are shown. For a spatially coherent electron source, the effect of the separation time $\Delta t = t_{pulse}$ between an electron pair and the coherence time $T_{coh}$ is considered. The vertical axis is the electron pulse duration or the detection time resolution, whichever is shorter. When the coherence time is larger than the time separation the quantum dip is complete (blue). For unpolarized electrons (Eq. 2), blue is defined as $2D_{HBT}^{rel}$. When the separation time between the electrons is shorter than about 20 fs, the classical dip is complete (red) and no coincidences are detected. No deviation from a random distribution, i.e., a dip of value one (black), is obtained for long separation times and short coherence times. A classically-caused dip is dominant for short time separations that exceed the coherence time. Classical Coulomb repulsion can thus mask the quantum degeneracy pressure (purple) when the separation time is short even if the coherence time is longer than the pulse duration. In the blue colored region, a pure quantum dip remains. The arrows point to the desired region of strong quantum degeneracy.

**STATISTICAL ANALYSIS.**

Electron pulses are photo-emitted with probability $P_n$, where the integer $n$ denotes the number of emitted electrons per incident laser pulse. For a purely random (Poissonian) emission process,

$$P_n(\lambda) = \frac{\lambda^n}{n!} e^{-\lambda}, \qquad (S2)$$

where the mean value $\lambda = r\tau_0$ and the emission rate is $r$. At an observed average detection rate of $1.5 \times 10^5$ s$^{-1}$ at each detector, the estimated emission rate is $10^7$ s$^{-1}$. (Even if the measured distribution is not Poissonian, estimates using a Poissonian distribution are accurate for the total countrate.) This gives $P_0 = 86.8\%$, implying $\lambda = 0.141$, with $P_1 = 12.3\%$, $P_2 = 0.864\%$, and $P_3 = 0.04\%$, etc. In other words, most laser pulses do not produce any electrons, while 93% of the pulses that do photoemit electrons produce one electron, and 6.5% produce two electrons. Therefore, the zero-time delay peak is dominated by $P_2$-events, while the surrounding peaks are dominated by $P_1$-events.

We demonstrated in the main text that a Coulomb model with reasonable parameters reproduces the dip $D$. Here, it is explained why the content of the zero-time delay peak must be compared with that of its neighboring tall peak (centered at $1\tau_0$) to judge whether the distribution is Poissonian or sub-Poissonian. We calculate the ratio of the number of counts in the zero-time delay peak, $N(0\tau_0)$, to the first neighboring peak, $N(1\tau_0)$. The total number of laser pulses during a single experiment is $N_p$. The probability for each emitted photoelectron to strike the start (stop) detector labeled A (B) is given by $\varepsilon_A$ ($\varepsilon_B$). The number of coincidence events, $N(m\tau_0)$, separated by a time interval $m\tau_0$, where $\tau_0$ is the repetition time as before and $m \geq 1$ is an integer, is given by

$$N(m\tau_0) = P_A \times \left(\tilde{P}_B\right)^{m-1} \times P_B \times (N_p - m), \qquad (S3)$$

where $P_A$ ($P_B$) is the probability to detect at least one electron in the detector A (B) over the time interval $\tau_0$, $\tilde{P}_B$ is the probability to detect no electrons in the stop detector during the same time interval, and the factor $(N_p - m)$ is the number of ways that one electron can be detected at A and another at B separated by $m\tau_0$ for $N_p$ laser pulses striking the nanotip.

The probability to detect an electron at the detector A and not have another electron trigger a stop signal at detector B is

$$P_A = \sum_{n=1}^{\infty} \sum_{n_A=1}^{n_A=n} P_n \binom{n}{n_A}\binom{n-n_A}{n_{\varepsilon 1}} \varepsilon_A^{n_A} \varepsilon_B^0 \varepsilon^{n_{\varepsilon 1}}, \qquad (S4)$$

where $P_n$ is given by Eq. (S2), the combinatorial $\binom{k}{p}$ is $\dfrac{k!}{p!(k-p)!}$, $n_{\varepsilon 1} + n_A = n$, and $\varepsilon = 1 - \varepsilon_A - \varepsilon_B$ is the probability to not be detected at either A or B, as the latter would close the coincidence window. In addition,

$$\begin{aligned}
\left(\tilde{P}_B\right)^{n-1} &= \left[P_0 + P_1(1-\varepsilon_B) + P_2(1-\varepsilon_B)^2 + \ldots + P_k(1-\varepsilon_B)^k + \ldots\right]^{n-1} \\
&= \left[\sum_{k=0}^{\infty} P_k(1-\varepsilon_B)^k\right]^{n-1} = \left\{\sum_{k=0}^{\infty} \frac{[\lambda(1-\varepsilon_B)]^k}{k!} e^{-\lambda}\right\}^{n-1} \\
&= e^{-(n-1)\lambda} \times e^{(n-1)\lambda(1-\varepsilon_B)} = e^{-(n-1)\lambda\varepsilon_B} = P_0^{(n-1)\varepsilon_B}.
\end{aligned} \qquad (S5)$$

The probability to detect an emitted electron at the detector B is

$$P_B = \sum_{n=1}^{\infty} \sum_{n_B=1}^{n_B=n} P_n \binom{n}{n_B}\binom{n-n_B}{n_{\varepsilon 2}} \varepsilon_B^{n_B} \varepsilon_h^{n_{\varepsilon 2}}. \qquad (S6)$$

Here, $n_{\varepsilon 2}$ equals $n - n_B$ as all the remaining electrons that are not detected in B land elsewhere with probability $\varepsilon_h = 1 - \varepsilon_B$.

To predict the number of coincidence events $N(m\tau_0)$ as a function of time delay $\tau = m\tau_0$, we consider its dependence on $m$. For a measuring time of 900 s, the number of laser pulses $N_p \sim 7 \times 10^{10} \gg m$ for $m < 10^4$, and the factor $(N_p - m)$ is nearly constant. The remaining $m$-dependent factor is $e^{-m\varepsilon_B \lambda} = e^{-m\varepsilon_B r \tau_0}$. For the zero-time delay peak, we find

$$N(0\tau_0) = N_p \times \sum_{n=2}^{\infty} \sum_{n_A=1}^{n_A=n-1} \sum_{n_B=1}^{n-n_A} P_n \binom{n}{n_A}\binom{n-n_A}{n_B}\binom{n-n_A-n_B}{n_{\varepsilon 3}} \varepsilon_A^{n_A} \varepsilon_B^{n_B} \varepsilon^{n_{\varepsilon 3}}, \tag{S7}$$

in which all the possible permutations for $n \geq 2$ emitted electrons to strike the detector pair are taken into account. Here, $n_{\varepsilon 3} = n - n_A - n_B$. To evaluate the ratio $N(0\tau_0)/N(1\tau_0)$, special cases of equations (S7) and (S3) are considered. For the first case, $\lambda \ll 1$ and $\varepsilon_i \ll 1$ with $i = A, B$, corresponding to our experimental parameters. Thus the Poissonian probabilities are a monotonically decreasing series and the ratio is given by

$$\frac{N(0\tau_0)}{N(1\tau_0)} \approx P_2 \left[\binom{2}{1}\binom{1}{1} \varepsilon_A^1 \varepsilon_B^1 \varepsilon^0 \right] / \left( P_1 \binom{1}{1} \varepsilon_A^1 \varepsilon_B^0 \varepsilon^0 \times e^{-\lambda \varepsilon_B (m-1)} \Big|_{m=1} \times P_1 \binom{1}{1} \varepsilon_B^1 \varepsilon_h^0 \right)$$

$$\approx \frac{\lambda^2 e^{-\lambda}}{2!} 2\varepsilon_A \varepsilon_B / \left( \frac{\lambda^1 e^{-\lambda}}{1!} \varepsilon_A \times \frac{\lambda^1 e^{-\lambda}}{1!} \varepsilon_B \right) = e^\lambda \approx 1 \tag{S8}$$

and is dominated by the first term in each summation. Consider instead $\lambda = 10$ and $\lambda \varepsilon_i \ll 1$, which corresponds to a source that emits about ten electrons per pulse. In this case, one finds $R_{01} = P(0\tau_0)/P(1\tau_0) = e^{\lambda \varepsilon_B}$. This demonstrates that the method is robust even for multi-electron experiments.

The emission rate determines $\lambda$ which sets $N(0\tau_0)/N(1\tau_0)$, allowing us to judge if the observed spectrum is sub-Poissonian. To quantify the amount of deviation from the Poissonian distribution (Fig. 2 main text), we define a statistical contrast $D/N^{\exp}(1\tau_0) = [N^{\exp}(0\tau_0) - N^{\exp}(1\tau_0)]/N^{\exp}(1\tau_0) = -24\%$, for which we have used the experimental values $N^{\exp}(0\tau_0) = 2.00 \times 10^5$ and $N^{\exp}(1\tau_0) = 2.64 \times 10^5$ (yellow and green regions in Fig. 2 of the main text). The above analysis is confirmed by a SIMION-based simulation of a pulsed emission process following a Poissonian distribution with an energy spread of 1 eV.

In the absence of Coulomb interaction between the electrons, near-equal heights are predicted for all the peaks in panel (a) (blue dashed line) and an exponential decay is predicted over the temporally stretched coincidence window of panel (b) (blue dashed line). For this window, the individual delay peaks are not resolved as the timing bin exceeds the repetition time.

1. Barwick, B. et al. Laser-induced ultrafast electron emission from a field emission tip. New J. Phys. **9**, 142 (2007).

2. https://simion.com/

3. Kiesel, H., Renz, A. & Hasselbach, F. Observation of Hanbury Brown-Twiss anticorrelations for free electrons. *Nature* **418**, 392-394 (2002).

4. Kodama, T., Osakabe, N. & Tonomura, A. Correlation in a coherent electron beam. *Phys. Rev. A* **83**, 063616 (2011).

5. Kuwahara M., et al. Intensity Interference in a Coherent Spin-Polarized Electron Beam, *Phys. Rev. Lett.* **126**, 125501 (2021).